\documentclass[baaa]{baaa}

\usepackage[pdftex]{hyperref}
\usepackage{subfigure}
\usepackage{natbib}
\usepackage{helvet,soul}
\usepackage[font=small]{caption}

\contriblanguage{1}

\contribtype{1}

\thematicarea{7}

\received{\ldots}
\accepted{\ldots}

\title{Unveiling the properties of the first galaxies\\with JWST and ALMA}

\titlerunning{First galaxies with JWST and ALMA}

\author{
M.E. De Rossi\inst{1}
\&
V. Bromm\inst{2}
}

\authorrunning{De Rossi et al.}

\contact{mariaemilia.dr@gmail.com}

\institute{
Instituto de Astronom{\'\i}a y F{\'\i}sica del Espacio, CONICET--UBA, Argentina
\and
Department of Astronomy, University of Texas at Austin, Estados Unidos
}

\resumen{
El {\sl Telescopio Espacial James Webb} (JWST) está desafiando nuestra comprensión de la naturaleza de las primeras galaxias del Universo, tras haber descubierto una sorprendente abundancia de galaxias masivas en épocas tempranas. Aplicando un modelo de polvo primigenio, estimamos las luminosidades en el infrarrojo lejano (FIR) para galaxias de diferentes masas a corrimientos al rojo $z\gtrsim7$. En particular, predecimos los flujos observados para diferentes bandas (3-10) del Atacama Large Millimeter/sub-mm Array (ALMA), considerando valores conservadores típicos para las propiedades de las primeras galaxias (por ejemplo, metalicidades gaseosas, relación polvo-metal, eficiencia de formación estelar). Como era de esperar, los flujos FIR aumentan con la masa estelar para todas las bandas de ALMA, pero con una pendiente mayor para las bandas 9 y 10. Es alentador que las primeras galaxias se vean afectadas por una fuerte corrección K negativa, de modo que las fuentes con propiedades similares son más brillantes en las bandas 3-8 a mayores corrimientos al rojo. Este comportamiento es más fuerte para las bandas 6-7. Aunque las tendencias para las bandas 9-10 no están claras, los flujos más altos para dichas bandas se alcanzan hacia el extremo $z\gtrsim 15$. Contraintuitivamente, nuestros resultados sugieren que las fuentes observadas por el JWST con similares masas y propiedades de polvo serían más fácilmente detectables con ALMA si se encuentran a $z$ mayores.
}

\abstract{
The {\sl James Webb Space Telescope} (JWST) is challenging our undestanding of the nature of the very first galaxies in the Universe, having discovered a surprising abundance of very massive galaxies in early cosmic epochs. By applying a model of primeval dust, we estimate the far-infrared (FIR)-continuum luminosities for galaxies of different masses at redshifts $z\gtrsim7$. In particular, we predict observed fluxes for different available bands (3-10) of the Atacama Large Millimeter/sub-mm Array (ALMA), considering typical conservative values expected for the properties of first galaxies (e.g., gas-phase metallicities, dust-to-metal ratio, star formation efficiency).  As expected, FIR fluxes increase with stellar mass for all ALMA bands, but with a steeper slope for bands 9 and 10.  Encouragingly, first galaxies are affected by a strong negative-K correction, in such a way that sources with similar properties are brighter in bands 3-8 at higher redshifts.  Such behaviour is stronger for bands 6-7.  Althought  the trends for bands 9-10 are not clear, the highest fluxes for such bands are reached towards extreme $z\gtrsim 15$.  Counterintuitively, our results suggest that JWST sources with similar masses and dust properties would be more easily detectable with ALMA if they are located at higher $z$.
}


\begin{document}

\maketitle
\section{Introduction}\label{sec:Introduction}
The study of the formation of the first cosmic structures is entering a golden epoch.  The advent of the {\sl James Webb Space Telescope} (JWST) has led to significant and exciting findings during the last years. In particular, different authors have reported a higher than expected abundance of massive galaxies at redshift $z \gtrsim 10$, whose origin and nature are still debated \citep[e.g.,][]{Finkelstein2022, Labbe2023, Adams2023}.
In order to obtain more insights into the nature of massive primeval sources discovered by the JWST, the synergy with other observational facilities will be crucial. In this sense, there have already been attempts to probe the dust continuum of such JWST sources with the Atacama Large Millimeter/submillimeter Array (ALMA) in the far-infrared (FIR) and submillimeter (sub-mm) bands \citep[e.g.,][]{Fujimoto2023}.

\begin{figure*}[!t]
\centering
\includegraphics[width=\columnwidth]{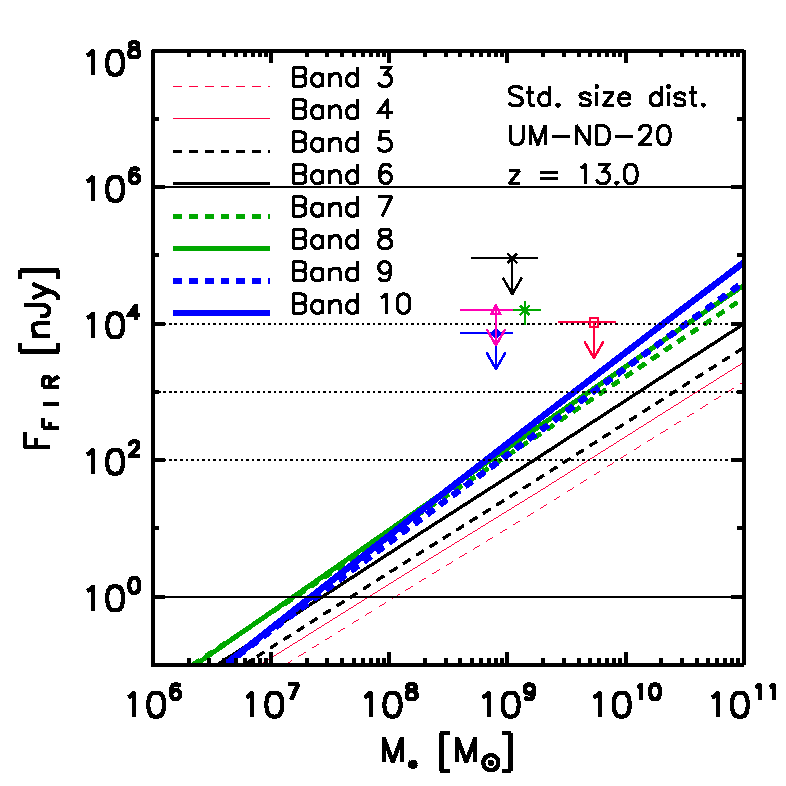}
\includegraphics[width=\columnwidth]{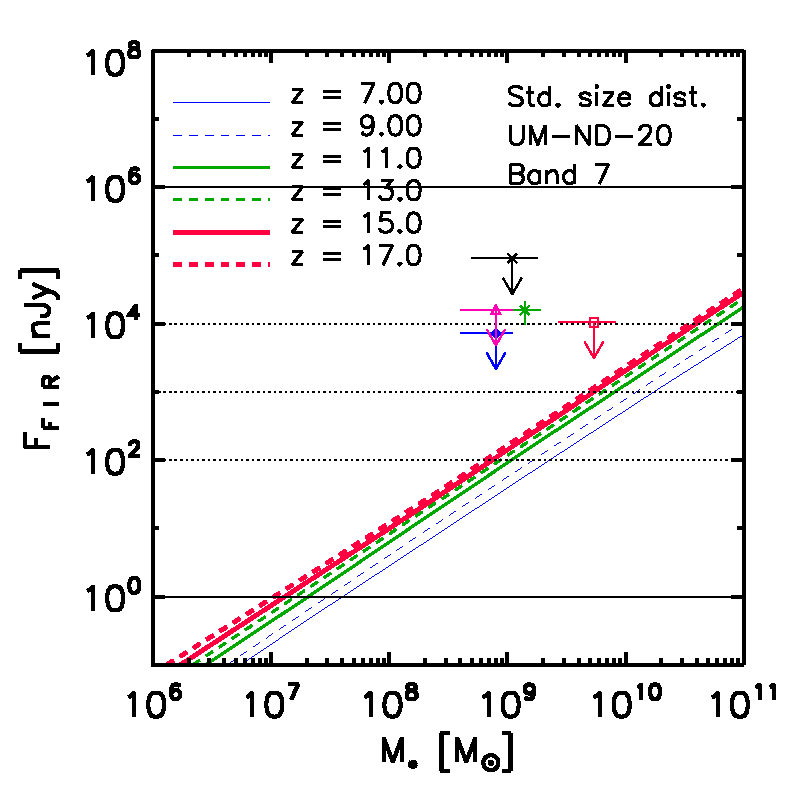}
\caption{
Average FIR flux as a function of $M_{*}$ for model galaxies. {\it Left panel}: Comparison between the fluxes obtained for different ALMA bands for a source at $z=13$.  {\it Right panel}: 
Comparison between the fluxes obtained for ALMA band 7 corresponding to similar sources located at different $z$.  Different symbols show upper limits (arrows) and a tentative value (green symbol) reported in the literature (see the text for details).}
\label{fig:fig1}
\end{figure*}

In \cite{DeRossi2023}, we implemented a model of primordial dust emission to predict the plausible signals of first galaxies which could be detected with ALMA in the FIR/sub-mm part of the spectrum.
According to our previous results, dust fractions and gas-phase metallicities ($Z_{\rm gas}$) in primeval galaxy sources can be constrained with upper FIR flux limits inferred from observations with ALMA. Encouragingly, when adopting model parameters (e.g., star formation efficiency -$\eta$-, $Z_{\rm gas}$, dust-to-metal mass ratio -$D/M$-) expected for typical first galaxies \citep[e.g.,][]{Jaacks2018,Jaacks2019}, our predicted FIR spectra are in agreement with upper flux limits derived from ALMA.  In particular, the observations rule out higher values of $Z_{\rm gas}$ ($\gtrsim 5\times 10^{-2}~ Z_{\odot}$) and $D/M$ ($\gtrsim 0.06$), unless a higher $\eta$ is assumed.

The analysis by \cite{DeRossi2023} focuses on galaxies at $z\sim 13$. For such systems, we evaluated the effects of varying model physical parameters on observed fluxes in ALMA band 7. We also predicted the possible fluxes resulting from multi-band observations of a galaxy of $M_\star \sim 10^9~\mathrm{M}_\odot$.  In this article, we extend our previous work, considering fixed model parameters and discussing the predictions of our model for wider ranges of $z$, $M_\star$, and ALMA bands.

\section{Dust model}
We implemented the dust emission model described in \citet{DeRossi2023}, which is in turn based on the methodology explained in detail in \citet{derossi2017} and \citet{derossi2019}.   Predictions of this model are suitable for the study of observed sources at high $z$ \citep[e.g.,][]{derossi2018}.  For the convenience of the reader, we here present a brief summary of our model; for further details, we refer to the aforementioned articles.

We assume that a model galaxy is composed of a central cluster of Population~II stars, surrounded by a mixed phase of gas and dust.  We consider this system to be located within a dark matter halo at $z \gtrsim 7$. In this work, we adopt model physical parameters corresponding to the so-called reference set in  \citet{derossi2017}: $D/M = 5 \times 10^{-3}$, $Z_{\rm g}=5\times10^{-3}~{\rm Z}_{\odot}$, and $\eta=0.01$.  These are typical conservative values expected for first galaxies (\citealt{greif2006, mitchellwynne2015, schneider2016}; see
\citealt{derossi2017, derossi2019}). 
Stellar radiation is obtained by using {\sc YGGDRASIL} model grids \citep[][]{zackrisson2011}.

Regarding the dust chemical composition, we consider the silicon-based UM-ND-20 chemistry described in \citet[][]{cherchneff2010}, which exhibits an intermediate behaviour with respect to the models studied by \citet{DeRossi2023}. We also adopt
the standard grain-size distribution implemented in \citet{ji2014}.
Following our previous works, we estimate the dust temperature ($T_{\rm d}$) by assuming thermal equilibrium and applying Kirchhoff’s law to determine the dust emissivity from the resulting $T_{\rm d}$ profile.

The observed dust specific flux $f_{\nu , {\rm obs}}$ from a model galaxy is estimated as:

\begin{equation}
f_{\nu , {\rm obs}} = (1 + z)  \frac{L_{\nu ,{\rm em}}}{4 \pi {d_{L}}^2}
\mbox{\, ,}
\end{equation}
where $L_{\nu ,{\rm em}}$ is the total specific dust luminosity and  $d_{L}$ is the luminosity distance to a galaxy at redshift $z$.

The average specific flux observed over a given ALMA band is given by:
\begin{equation}
\label{eq:average_flux}
F_{\rm FIR} = \frac{\int_{{\nu}_{i}}^{{\nu}_{f}} f_{\nu , {\rm obs}} \ {\rm d}{{\nu}_{\rm obs}}}{{{\nu}_{f}}-{{\nu}_{i}}},
\end{equation}
where ${{\nu}_{i}}$ and ${{\nu}_{f}}$ indicate the frequency range corresponding to
that band. We evaluate the same bands as in \citet{DeRossi2023} (Band 3: 84-116 GHz, Band 4: 125-163 GHz, 
Band 5: 163-211 GHz, Band 6: 211-275 GHz, Band 7: 275-373 GHz, Band 8: 385-500 GHz, 
Band 9: 602-720 GHz, Band 10: 787-950 GHz).

Finally, we adopt a $\Lambda$ Cold Dark Matter ($\Lambda$CDM) cosmology with $h$ = 0.67,
${\Omega}_{\rm b}$ = 0.049,
${\Omega}_{\rm m}$ = 0.32,
${\Omega}_{\Lambda}$ = 0.68 \citep{Planck2014}.

\begin{figure*}[!t]
\centering
\includegraphics[width=\columnwidth]{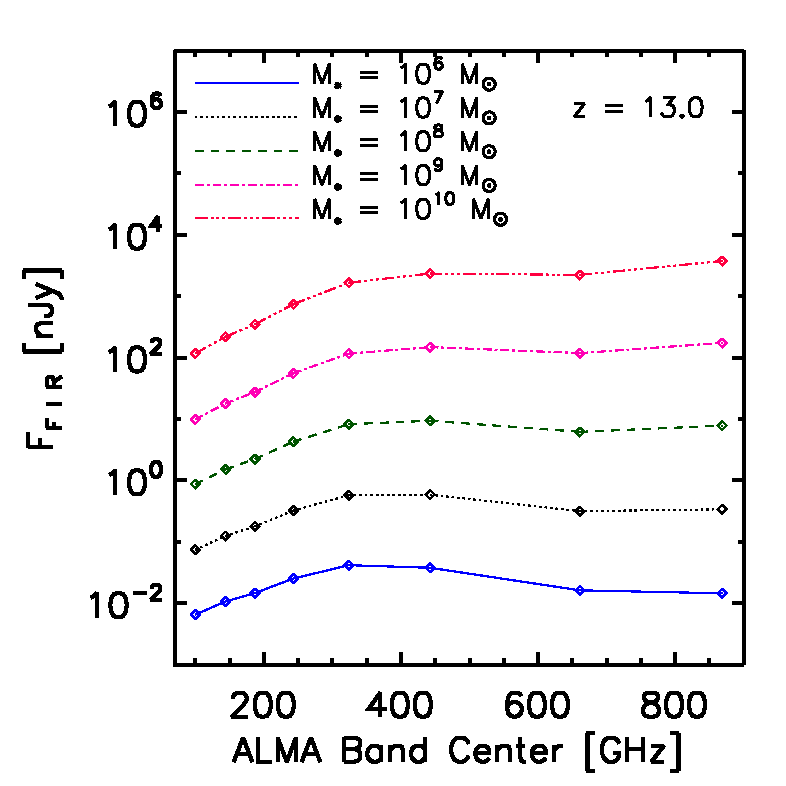}
\includegraphics[width=\columnwidth]{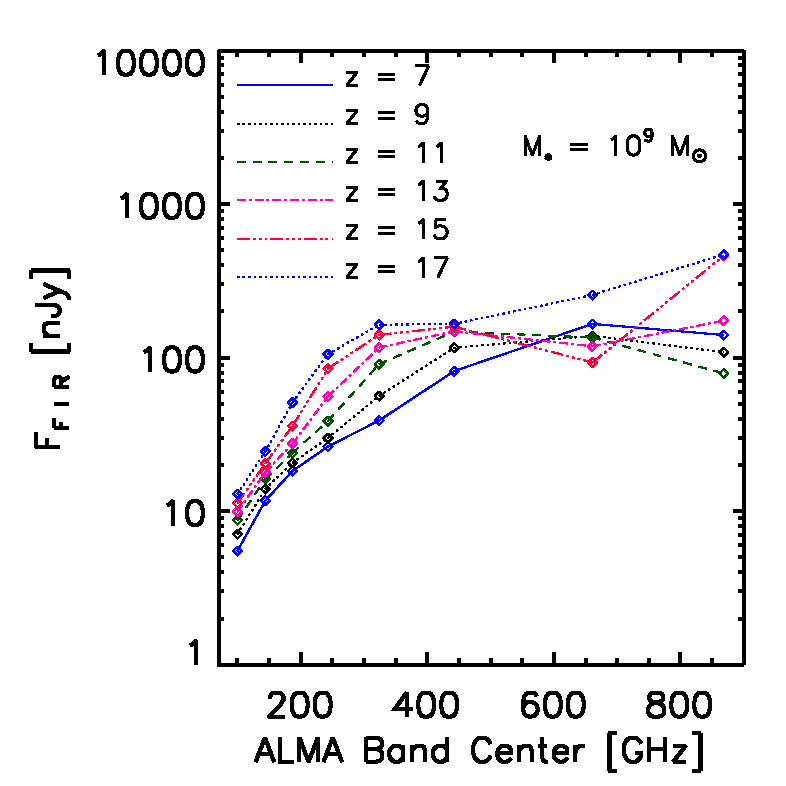}
\caption{
 Average FIR flux within available ALMA bands (3\--10) vs. central band frequency.
 {\it Left panel}:  Comparison of the predictions obtained for sources of different $M_*$ located at $z=13$.   {\it Right panel}:  Comparison of the predictions obtained for sources of $M_* = 10^{9} ~ {\rm M}_{\odot}$ located at different $z$.
}
\label{fig:fig2}
\end{figure*}

\section{Results}

In Fig.~\ref{fig:fig1} (left panel), we evaluate $F_{\rm FIR}$ as a function of $M_*$ for model sources at $z=13$, considering different ALMA bands.  We obtain a linear increase of FIR fluxes with $M_*$ for all analysed bands, with a steeper slope and normalisation for bands associated with higher frequencies.  In particular, sources with higher $M_*$ show the largest variations of FIR fluxes among different bands, reaching a difference of $\sim 2$ orders of magnitude between the fluxes corresponding to ALMA bands 3 and 10 at $M_\star \sim 10^{11}~\mathrm{M}_\odot$.
Different symbols in Fig.~\ref{fig:fig1} denote ALMA observational data regarding select galaxy candidates at $z\gtrsim 11$  \citep[][see their tables 1 and 3]{Fujimoto2023}, corresponding to sources:
S5-z17-1 (band 7, \citealt{Fujimoto2023}; black cross), GHZ1/GLz11 (band 7, \citealt{Yoon2023}; green asterisk), GHZ2/GLz13 (band 6, \citealt{Bakx2023}, \citealt{Popping2023}; blue circle), and HD1 (band 6, \citealt{Harikane2022}, pink triangle; band 4, \citealt{Kaasinen2023}, red square), plausibly located at $z\approx 18.41$, $10.87$, $12.43$, and $15.39$, respectively \citep{Fujimoto2023}. 
A tentative detection is represented for GHZ1/GLz11, while data for all other sources correspond to upper flux limits.
As noted by \citet{DeRossi2023}, predictions of our model are consistent with upper flux limits inferred from ALMA, as long as conservative typical parameter values for the baryonic properties in first galaxies are assumed.

In Fig.~\ref{fig:fig1} (right panel), we analyse the predictions of our model regarding the evolution with $z$ of ALMA fluxes in band 7, which is commonly used in observational studies.  Interestingly, systems with similar $M_*$ would be brighter in the FIR if they are located at higher $z$. In particular, as $z$ increases from 7 to 17, the fluxes increase by around $\sim 1$ order of magnitude for all $M_*$ considered. \cite{derossi2019} obtained similar trends when evaluating the capabilities of a projected future space telescope for detecting first galaxies in the FIR.  According to that work, first galaxies are affected by a strong negative K-correction. On the one hand, the higher temperature floor set by the cosmic microwave background drives higher dust luminosities as $z$ increases \citep[e.g.,][]{Safranek-Shrader2016}.
On the other hand, stellar populations can more efficiently heat dust grains at higher $z$ as systems are more spatially concentrated.
Hence, considering galaxy candidates of similar $M_*$ observed by the JWST, galaxies at higher $z$ would be more amenable for detection with ALMA. But, this would be valid as long as dust properties do not vary significantly with $z$, which is currently not clear.

Fig.~\ref{fig:fig2} (left panel) presents predictions of our model corresponding to multi-band observations conducted with ALMA for sources of different $M_*$ at $z=13$. The shape of the relation between $F_{\rm FIR}$ and the central band frequency does not seem to significantly depend on $M_*$: $F_{\rm FIR}$ shows a weak increase with the central frequency for bands 3-7 and, then, the relation flattens for bands 7-10.  Nevertheless, for the latter bands, we note a very weak decrease of the slope for less massive galaxies.  In particular, the increase of $F_{\rm FIR}$ with $M_*$ is larger at higher frequencies, as Fig.~\ref{fig:fig2} also shows:  as $M_*$ increases in the range $10^{6-10}~\mathrm{M}_\odot$, fluxes in band 3 increase by $\sim 4$ orders of magnitude, while fluxes in band 10 increase by $\sim 6$ orders of magnitude.

Multi-band observations with ALMA for sources of $M_* = 10^{9}~\mathrm{M}_\odot $ at different $z$ are analysed in  Fig.~\ref{fig:fig2} (right panel). It is evident that, for sources of similar $M_*$, the shape of the relation between $F_{\rm FIR}$ and the central band frequency is affected by $z$.  This can be explained considering that different features of the dust spectral energy distribution enter the wavelenght range of each band as $z$ increases (see the discussion in \citealt{derossi2019}). Such spectral features are generated by the specific chemical composition of dust grains.  Therefore, ALMA multi-band observations could provide information about the chemistry of dust as a function of $z$, allowing us to constrain its evolution.
We note again that galaxies at the highest redshifts ($z\sim17$) tend to be brighter in all bands compared with similar systems at $z=7$, due to the strong negative K-correction already mentioned. In general, $F_{\rm FIR}$ increases with $z$ for bands 3-8, reaching the highest variation for bands 6-7.  The trends with $z$ are not so clear in the case of bands 9-10, but the highest fluxes are obtained for such bands at our maximum evaluated value of $z=17$.  This strong negative K-correction obtained for first galaxies can thus facilitate the exploration of the origin and nature of dust in the early Universe \citep[e.g.,][]{Gall2011}.

\section{Conclusions}
By implementing a model for primordial dust emission, we studied the plausible FIR/sub-mm signals of first galaxies at $z \gtrsim 7$.  We predicted the observed fluxes corresponding to different ALMA bands, considering sources within a stellar mass range $M_* \sim 10^{6-11}~\mathrm{M}_\odot$. 
In agreement with previous works, our results are consistent with current upper flux limits from ALMA.

As expected, FIR fluxes increase nearly proportionally to the mass of the systems, but with a slope depending on the ALMA band considered.
Contrary to expectations, fluxes corresponding to ALMA bands 3-8 are higher as $z$ increases; thus, sources with similar masses are brighter in the FIR for larger cosmological distance. This is a consequence of the strong negative K-correction that affects galaxies at high $z$, as discussed in detail in \citet{derossi2019}. Therefore, if the dust properties do not change considerably with $z$, the JWST sources of a given $M_*$ could be detected more easily with ALMA if they are located at higher $z$.

Our findings support the great potential provided by the synergy between the JWST and ALMA to address the nature of dust in the early Universe, which in turn is a powerful probe of cosmic chemical evolution in the formative first billion years \citep[e.g.][]{Karlsson2013}.

\begin{acknowledgement}
We thank Alexander Ji for providing tabulated dust opacities for the different
dust models used here.
This work makes use of the Yggdrasil code \citep{zackrisson2011}, which adopts
Starburst99 SSP models, based on Padova-AGB tracks \citep{leitherer1999, vazquez2005}
for Pop~II stars. We acknowledge support from {\it Agencia Nacional de Promoci\'on de la Investigaci\'on, el Desarrollo Tecnol\'ogico y la Innovaci\'on} (Agencia I+D+i, PICT-2021-GRF-TI-00290, Argentina). 

\end{acknowledgement}


\bibliographystyle{baaa}
\small
\bibliography{bibliografia}
 
\end{document}